\documentclass[aps, pra, notitlepage]{revtex4-1}
\usepackage[dvipdf]{epsfig}
\usepackage{color}
\usepackage{color}
\usepackage{amsmath,amssymb,mathrsfs}
\def\vf{{\epsilon^\star_f }}

\def\q{{ Q}^\star}
\def\bc{{\bf c }}
\def\bx{{\bf x }}
\def\bg{{\bf g }}
\def\bk{{\bf k }}
\def\a{{\alpha }}
\def\b{{\beta }}

\def\A{{\cal{A}^\star }}

\def\Ef{{{\rm E_{1}}(\vf)}}

\def\d{\textsf{d}}

\def\sf{\textsf{s}_\textsf{f}}
\def\sr{\textsf{s}_\textsf{r}}

\def\s{\textsf{s}}

\def\aa{\textsf{a}}
\def\be{\begin{equation}}
\def\ee#1{\label{#1}\end{equation}}

\def\s{\textsf{s}}
\def\kk{\textsf{k}}
\newcommand{\ben}{\begin{eqnarray}}
\newcommand{\een}{\end{eqnarray}}
\begin{document}
\title{Analysis of the Reaction Rate Coefficients for Slow
  Bimolecular Chemical Reactions}
\thanks{Dedicated to Professor
    I-Shih Liu on the occasion of his 70th birthday.}
\author{Gilberto M. Kremer}
\email{kremer@fisica.ufpr.br}
\author{Tiago G.  Silva}\email{tgutierres@fisica.ufpr.br}
\affiliation{Departamento de F\'{\i}sica, Universidade Federal do Paran\'a, Curitiba, Brazil}
\pacs{51.10.+y (Kinetic and transport theory of gases);      47.70.Fw (Chemically reactive flows)    }
\keywords{Boltzmann equation, Chemically reactive flows, Reaction rates}

\begin{abstract}
  Simple bimolecular reactions $A_1+A_2\rightleftharpoons A_3+A_4$ are
  analyzed within the framework of the Boltzmann equation in the
  initial stage of a chemical reaction with the system far from
  chemical equilibrium. The Chapman-Enskog methodology is applied to
  determine the coefficients of the expansion of the distribution
  functions in terms of Sonine polynomials for peculiar molecular
  velocities. The results are applied to the reaction $H_2
  +Cl\rightleftharpoons HCl+H$, and the influence of the
  non-Maxwellian distribution and of the activation-energy dependent
  reactive cross sections upon the forward and reverse reaction rate
  coefficients are discussed.
\end{abstract}
\maketitle

\section{Introduction}
\label{sec:1}
Since the works of Prigogine and collaborators in the 1950's, the
study of chemically reacting gases by means of the Boltzmann equation
has constituted a topic of research \cite{PX,PM}.  The first article
in this series analyzed the chemical reaction $A+A\longrightarrow
B+C$ under the assumption that the constituents have only
translational energy and the reagents are more concentrated than the
products \cite{PX}. The Boltzmann equation was solved by the Chapman-Enskog
method, the distribution function was expanded to second order in the
Sonine polynomials, and the reaction rate was determined for two kinds
of reactive differential cross sections: one of them a step
function, the other taking the activation energy into account.

The second work analyzed the chemical reaction
$A_0+B\rightleftharpoons A_1+B$ and gave attention to the reaction
heat \cite{PM}. The dependence of the reaction rate on the activation
energy and on the reaction heat were determined.

It has long been known that the reaction rate depends on the choice of
the molecular interactions, and this motivated Present to propose the
following expression for the reactive cross section \cite{1,2}
\begin{align}\label{aa}
\sigma^\star = \left\{
  \begin{array}{ll}
    0, &E\leq\epsilon^\star, \\
    \pi D^2\left[1-(\epsilon^\star/E)  \right],&E>\epsilon^\star.
  \end{array}\right.,
\end{align} where the formation of the activation complex is possible
only when the distance between the molecular centers is equal to the parameter
$D$, related to the diameters of the molecules. This cross section, known as the \emph{line-of-centers
model}, allows a chemical reaction whenever the relative
translational energy of the molecules $E$ is larger than the
activation energy $\epsilon^\star$.

In 1960 Ross and Mazur \cite{3} analyzed the bimolecular chemical
reaction $A+B\rightleftharpoons C+D$ and extracted general expressions
for the forward  and reverse reaction rates and the entropy production
from the Boltzmann equation. The distribution function was not
determined explicitly from the Boltzmann equation so that no exact
expressions for the reaction rates and entropy production were
derived.  The solution of the Boltzmann equation for the distribution
function from the expansion of the distribution function in
Sonine polynomials and computation of the forward and reverse reaction rate were
presented later by Present \cite{4} and by Shizgal and Karplus \cite{5,6,7,8}.

Many contributions to the analysis of chemically reactive systems
follow these works. Among others we quote
\cite{9,10,11,12,13,14,15,16,17,18,19,20,21,22,23,24,25,26,27,28,29,30}.

The cross sections determining the reactive collision term can be
divided into two types, namely with and without activation energy
\cite{St,GK}. In general, cross sections with activation energy allow
a reaction to occur whenever the relative translational energy exceeds
the activation energy, as in Eq.~(\ref{aa}). Even the slightest
grazing collision leads to a chemical reaction. In a more realistic
scenario, a reaction occurs only when the relative translational
energy in the direction of the line joining the centers of the
molecules is larger than the activation energy \cite{9,29,GK}. In this
case the geometry of the collision plays a fundamental role in
controlling the occurrence of a reaction.

Another important issue in applications of the Chapman-Enskog to the
reactive Boltzmann equation is the distortion of the Maxwellian
distribution functions by the reaction heat \cite{PM,12,24,29}, which
makes the reaction rate coefficients sensitive to the reaction heat.

The aim of this work is to analyze the influence of the
activation-energy cross sections and of the non-Mawellian
distribution function on the reaction-rate coefficients. Although
relying on the same method adopted in the above-mentioned works, we
here solve the Boltzmann equation for the distribution function, which
we expand in Sonine polynomials up to second order. The reactive
differential cross section we consider allows reactions only when the
relative translational energy in the direction of the line joining the
centers of the molecules is greater than the activation energy, and we
analyze slow reactions. In other words, we analyze the initial stage
of a chemical reaction, in which the system is far from chemical
equilibrium. In this stage the elastic collisions are more frequent
than reactive ones, and the affinity is much larger than the thermal
energy of the mixture. We show that the reactive cross section under
study markedly influences the reaction rates, while the effect of the
non-Maxwellian distribution is not too pronounced.

The work is structured as follows: Section 2 introduces the system of
Boltzmann equations for a simple bimolecular chemical reaction
$A_1+A_2\rightleftharpoons A_3+A_4$. The expressions for the reaction
rate coefficients that follow from the Boltzmann equation and the
specification of the elastic and reactive cross sections are the
subject of Section 3. In Section 4 the Arrhenius equation is obtained
from a Maxwellian distribution function and it is shown that the
reaction-rate coefficient when the relative translational energy in
the direction of the line which joins the centers of the molecules
must be larger than the activation energy is different from the
coefficient when only the relative translational energy must be
greater than the activation energy. The analysis of the slow chemical
reactions is the subject of Section 5, where the
distribution functions are expanded in Sonine polynomials of the
peculiar molecular velocity and the first coefficients of the
expansions are obtained with the Chapman-Enskog methodology. In
Section 6 the results of the previous section are applied to the chemical
reaction $H_2 +Cl\rightleftharpoons HCl+H$, and the coefficients of
the non-Maxwellian distribution function, and the forward and reverse
reaction coefficients are plotted as functions of the temperature. The
results show the effect of the non-Maxwellian distribution
functions and of the reactive cross sections on the reaction-rate
coefficients and on the entropy production rate.

\section{Boltzmann Equations}
\label{sec:2}
We consider a simple reversible bimolecular gas reaction characterized
by the chemical law $A_1+A_2\rightleftharpoons A_3+A_4$, which takes
elastic and reactive binary encounters between the molecules into
account.

The elastic collisions between the two constituent molecules, of
masses $m_\alpha$ and $m_\beta$, have asymptotic pre-collisional
velocities $(\bf c_{\alpha}, \bf c_{\beta})$, asymptotic
post-collisional velocities $(\bf c_{\alpha} ', \bf c_{\beta} ')$, and
asymptotic relative velocities ${\bf g}_{{\beta} {\alpha}}={\bf
  c}_{\beta}-{\bf c}_{\alpha}$ and ${\bf g}_{{\beta} {\alpha}}'={\bf
  c}_{\beta}'-{\bf c}_{\alpha}'$, respectively, so that the
conservation laws of linear momentum and energy are given by the expressions
\begin{align}\label{1.84b} m_{\alpha} {\bf c}_{\alpha} + m_{\beta} {\bf
c}_{\beta}= m_{\alpha} {\bf c}_{\alpha}' + m_{\beta} {\bf
c}_{\beta}', \\\label{1.84c} {1 \over 2} m_{\alpha} c_{\alpha}^2 + {1
\over 2} m_{\beta} c_{\beta}^2= {1 \over 2} m_{\alpha}
c_{\alpha}^{\prime 2}+ {1 \over 2} m_{\beta} c_{\beta}^{\prime 2},
\end{align} respectively, where $\alpha, \beta=1,\dots,4$. The energy
conservation law can also be written in terms of the asymptotic
relative velocities as $g_{{\beta} {\alpha}}=g'_{{\beta} {\alpha}}.$

 For a reactive collision the conservation laws of mass, linear
 momentum and total energy read \begin{align}\label{9.27} m_1+m_2 = m_3+m_4,
 \\\label{9.28} m_1{\bc}_1 + m_2{\bc}_2 = m_3{\bc}_3 +
 m_4{\bc}_4,\\\label{9.29} \epsilon_1+\frac{1}{2}m_1
 c_1^2+\epsilon_2+\frac{1}{2}m_2 c_2^2= \epsilon_3+\frac{1}{2}m_3
 c_3^{2}+\epsilon_4+\frac{1}{2}m_4 c_4^{2},  \end{align}
respectively. Here $\epsilon_\alpha $ denotes the molecular binding energy
($\a=1,\dots,4$), while $(\bc_1, \bc_2)$ are the
 velocities of the reactants, and $(\bc_3, \bc_4)$, the velocities of
 the products of the forward reaction. The conservation law for the
 total energy (\ref{9.29}) can be written in terms of the relative
 velocities  $g_{\b\a}=\vert{\bg}_{\b\a}\vert=\vert{\bc}_\b-{\bc}_\a\vert$ and of
 the heat of reaction $Q$\textemdash defined as the difference of the
 chemical binding energies of the products and the reactants
 $Q=\epsilon_3+\epsilon_4-\epsilon_1-\epsilon_2$\textemdash as
\be
 \frac{1}{2}m_{12}g_{21}^2 = \frac{1}{2} m_{34}g_{43}^{ 2}+Q,
 \ee{9.30}
where $m_{\a\b}=m_\a m_\b/(m_\a+m_\b)$ denotes the reduced
 mass.

 We characterize the state of a reacting gaseous mixture in the phase
 space spanned by the positions $\bx$ and velocities ${\bc}_\a$ of the
 molecules by the set of distribution functions $f_\alpha \equiv
 f(\bx, \bc_\alpha ,t)$ with $\alpha=1,\dots,4$.  The distribution
 function $f_\a$ is defined so that the number of $\alpha$ molecules
 in the volume element $d{\bx}d{\bc}_\alpha$ around the position
 ${\bx}$ and the velocity ${\bc}_\alpha $ at time $t$ is given by
 $f_\alpha d{\bx}d{\bc}_\alpha$.

 The phase-space evolution of the distribution function $f_\alpha $
 for constituent $\a$ is governed by the Boltzmann equation, which in
 the absence of external forces reads
\be
 \frac{\partial{f}_\alpha}{\partial{t}} +
 c_{i}^{\alpha}\frac{\partial{f}_\alpha}{\partial{x}_i} =
 \sum_{\beta=1}^4\int\left(f^\prime_\alpha f^\prime_\beta-f_\alpha
   f_\beta\right) g_{\beta\alpha}\sigma_{\beta\alpha}
 \,\textrm{d}\Omega\,\textrm{d}{\mathbf{c_\beta}}+ {\cal Q}_\alpha^R,\qquad \a=1,\dots,4,
 \ee{9.31}
with the shorthand $f_\a^\prime\equiv f_\a(\bx,\bc_\a^\prime,t)$.

The left-hand side of Eq.~\eqref{9.31} refers to the
 space-time evolution of the distribution function, while its
 right-hand side takes the molecular collisions into account.
 The latter has two terms. The first one describes the elastic
 interactions among the constituents. In the integrand, the factor
 $\sigma_{\alpha \beta }$ is the differential elastic cross section,
 and $\mathrm{d}\Omega=\sin \chi d\chi d\varepsilon$ the solid-angle
 element, with $\chi$ denoting the scattering angle, and $\varepsilon$, the
 azimuthal angle that characterizes the collision.

 The second term ${\cal Q}_\alpha^R$ on the right-hand side of
 Eq.~(\ref{9.31}) is related to the reactive collisions. The
 expression for the constituent labeled by the index 1 is obtained as
 follows. The number of reactive collisions for the forward reaction
 $A_1+A_2\rightharpoonup A_3+A_4$ per unit of volume and time is given
 by $(f_1\, f_2\, g_{21} \,\sigma_{12}^\star \,d\Omega^\star \,d{\bf
   c}_1 \, d{\bf c}_2),$ where $\sigma_{12}^\star$ represents the
 reactive differential cross section and $\mathrm{d}\Omega^\star$ is
 the solid-angle element relating the orientation of the
 post-collisional relative velocity ${\bf g}_{43}$ with respect to the
 pre-collisional one ${\bf g}_{21}$. Likewise, the number of
 collisions for the reverse reaction $A_1+A_2\leftharpoondown A_3+A_4$
 reads $(f_3\, f_4\, g_{43} \,\sigma_{34}^\star \,d\Omega^\star
 \,d{\bf c}_3 \, d{\bf c}_4).$ For a fixed value of the heat of
 reaction it follows from Eq.~(\ref{9.30}) that
 $m_{12}g_{21}\,\mathrm{d}g_{21} = m_{34}g_{43}\,\mathrm{d}g_{43}$,
 and we also may obtain the equality $ m_{34}g_{21} \,\mathrm{d}{\bf
   c}_3\,d{\bf c}_4={m_{12}{g}_{43}}\,\,\mathrm{d}{\bf
   c}_1\,\mathrm{d}{\bf c}_2$.  Furthermore, for small
 heats of reaction the
 principle of microscopic reversibility yields
 $(m_{12}g_{21})^2\sigma^\star_{12}=(m_{34}g_{43})^2\sigma^\star_{34}$.
 Hence, all relationships accounted for, we find that
 $f_3\, f_4\, g_{43} \,\sigma_{34}^\star \,d\Omega^\star \,d{\bf c}_3
 \, d{\bf c}_4=f_3\, f_4\, \left( {m_{12}}/{m_{34}}\right)^3\, g_{21}
 \,\sigma_{12}^\star \,\mathrm{d}\Omega^\star \,\mathrm{d}{\bf c}_1 \, \mathrm{d}{\bf c}_2$. From
 the difference of the expressions for the number of collisions in the
 backward and forward reactions divided by $\mathrm{d}\bc_1$ and integrated
 over all values of $\mathrm{d}\bc_2$ and $\mathrm{d}\Omega^\star$, we
 obtain the following expression for the
 reactive collision term for the constituent labeled 1:
\be {\cal Q}^R_{1(2)} = \int\left[f_3 f_4\left(
     \frac{m_{12}}{m_{34}}\right)^3-f_1 f_2\right]\sigma_{12}^\star
 g_{21}d\Omega^\star d{\bf c}_{2(1)}.  \ee{9.35}

 The reactive collision term for the constituent labeled by the index
 2 being similar, the terms within parentheses in Eq.~\eqref{9.35}
 determine ${\cal Q}^{R}_{2}$. In the same notation, the reactive
 collision terms for the constituents labeled by the indexes 3 and 4
 read \be {\cal Q}^R_{3(4)} = \int\left[f_1
   f_2\left(\frac{m_{34}}{m_{12}}\right)^3-f_3
   f_4\right]\sigma_{34}^\star g_{43} \textrm{d}\Omega^\star \textrm{d}{\bc}_{4(3)}.
 \ee{9.36}

\section{Reaction Rates and Differential Cross Sections}
\label{sec:3}

The description of the evolution equation for the particle number
density $n_\a=\int f_{\alpha}\,d{\bf c}_{\alpha}$ of constituent $\a$
is obtained by integrating the Boltzmann equation (\ref{9.31}) over
all values of ${\bc}_\alpha$, which yields \be {\partial
n_{\alpha}\over \partial t} +{\partial
n_{\alpha}v_i^{\alpha}\over \partial x_i}=\tau_\a.  \ee{9.41}
Here, the bulk velocity $v_i^\a$ and the particle number density
production $\tau_\a$ of the constituent $\a$ read \be
v_i^\a=\frac{1}{n_\a}\int c_i^\a f_\a d\bc_\a,\qquad \tau_\a = -
\nu_\a \left(n_3 n_4\kk_r-n_1 n_2\kk_f\right), \ee{9.43}
where $\kk_f$ and $\kk_r$ denote the forward and reverse reaction rate
coefficients, respectively. They are defined by
 \begin{align}\label{9.44} \kk_f={1\over n_1n_2}\int f_1 f_2 \sigma_{12}^\star
 g_{21} \mathrm{d}\Omega^\star \mathrm{d}{\bf c}_{1} d{\bf c}_{2},\qquad \kk_r={1\over
 n_3n_4}\int f_3 f_4 \left( \frac{m_{12}}{m_{34}} \right)^{3}
 \sigma_{12}^\star g_{21} \mathrm{d}\Omega^\star \mathrm{d}{\bf c}_{1} \mathrm{d}{\bf c}_{2}.
\end{align} In Eq.~(\ref{9.43})$_2$ we have introduced the stoichiometric
coefficients $\nu_\a$ of the constituent $\a$, which for the chemical
reaction $A_1+A_2\rightleftharpoons A_3+A_4$ are given by
$\nu_1=\nu_2=-\nu_3=-\nu_4=-1$.

In order to determine the distribution functions $f_\a$ from the
system of Boltzmann equations~(\ref{9.31}), we have to specify the
elastic and the reactive differential cross sections. We assume that
the elastic differential cross sections correspond to a hard-sphere
potential, i.~e., that
 \be \sigma_{\alpha \beta } = {1\over4}\d_{\alpha \beta }^2,\qquad
\d_{\alpha \beta } = {1\over2}\left(\d_\alpha+\d_\beta\right),
 \ee{9.56} where $\d_\alpha$ and $\d_\beta$ represent the diameters of
 the colliding spheres.

For the reactive differential cross section we shall use the modified
line-of-centers model \cite{9,29,GK}
\begin{subequations}
  \begin{align}\label{9.58a} \sigma_{12}^\star = \left\{
      \begin{array}{ll}0,
        &{m_{12}\left(\bk_{21}\cdot\bg_{21}\right)^2/2}<\epsilon_f,
        \\
        {1\over4}\sf\d_{12}^2\left[1-{2\epsilon_f\over
            {m_{12}\left(\bk_{21}\cdot\bg_{21}\right)^2}}
        \right],&{m_{12}\left(\bk_{21}\cdot\bg_{21}\right)^2/2}\geq\epsilon_f,
      \end{array} \right.  \\
    \label{9.58b}
    \sigma_{34}^\star = \left\{
      \begin{array}{ll}
        0,
        &{m_{34}\left(\bk_{43}\cdot\bg_{43}\right)^2/2}<\epsilon_r,
        \\
        {1\over4}\sr\d_{34}^2\left[1-{2\epsilon_r\over
            {m_{34}\left(\bk_{43}\cdot\bg_{43}\right)^2}}
        \right],&{m_{34}\left(\bk_{43}\cdot\bg_{43}\right)^2/2}\geq\epsilon_r.
      \end{array} \right.
  \end{align}
\end{subequations}
Here $\epsilon_f$ and $\epsilon_r$ denote the
forward and reverse activation energies, respectively, and $\sf$ and
$\sr$ the corresponding steric factors, while $\bk_{21}$ and
$\bk_{43}$ are unit collision vectors in the directions of the centers
of the colliding molecules pointing from the centers of 2 and 4 to the
centers of 1 and 3, respectively.

The differential cross sections in Eqs.~\eqref{9.58a}~and
\eqref{9.58b} take into account the activation energies and the
geometry of the reactive collisions, since they are functions of the
relative translational energies in the direction of the lines joining
the centers of the molecules
${m_{12}\left(\bk_{21}\cdot\bg_{21}\right)^2}$ and
${m_{34}\left(\bk_{43}\cdot\bg_{43}\right)^2}$.  The original
line-of-centers model considers the relative translational energies as
${m_{12}g_{21}^2}$ and ${m_{34}g_{43}^2}$, so that even the slightest
grazing collision leads to a chemical reaction.

\section{Arrhenius Equation}
\label{sec:4}
For mixtures the chemical potential of the constituent $\a$, which takes
into account the binding energy of the molecules\textemdash but not the
internal states of the molecules associated with the rotational,
vibrational, electronic and nuclei states\textemdash is given by the
equality
\be
\mu_\a=\frac{kT}{m_\a}\left[\frac{\epsilon_\a}{kT}+\ln
n_\a-\frac{3}{2}\ln T+\mathcal{C}_\a\right],  \ee{a0} where $k$
denotes Boltzmann's constant, $\mathcal{C}_\a$ is a constant, and
all constituents have been assumed to be at the temperature $T$ of the mixture,
defined by
\be T=\frac{1}{3kn_\a}\int
m_\a \vert \bc_\a-{\bf v}_\a\vert^2 f_\a d\bc_\a.  \ee{temp}

Chemical equilibrium is characterized by the
condition
\be \sum_{\a=1}^4 m_\a \nu_\a \mu_\a^{\rm eq}=0, \ee{a1}
where the index "eq" denotes the equilibrium-value of the chemical
potential. Out-of-equilibrium chemical reactions are described by the
affinity, defined as
\be \mathcal{A}=-\sum_{\a=1}^4 m_\a \nu_\a \mu_\a,
\qquad\hbox{with} \qquad \mathcal{A}^{\rm eq}=0.  \ee{a2}

From the above equations we can obtain the law of mass action,
\be
\ln\left[\frac{n_1^{\rm eq}n_2^{\rm eq}}{n_3^{\rm eq}n_4^{\rm
eq}}\left(\frac{m_3m_4}{m_1m_2}\right)^{\frac{3}{2}}\right]=\q,
\ee{a3}
and the following expression for the affinity
\be
\ln\left[\frac{n_1n_2n_3^{\rm eq}n_4^{\rm eq}}{n_3n_4n_1^{\rm
eq}n_2^{\rm eq}}\right]=\A, \ee{a4}
where we have introduced the dimensionless reaction heat $\q=Q/kT$ and
affinity $\A=\mathcal{A}/kT$, in units of the thermal energy $kT$.

The state of equilibrium of a non-reacting gas mixture is
characterized by the Maxwellian distribution functions
\be f^{
(0)}_{\alpha}=n_{\alpha} \left ({m_\a\over2\pi kT}\right)^{3\over
2}e^{-\frac{m_\a \xi_{\alpha}^2}{2kT}},\qquad \alpha=1,\dots, 4,
\ee{9.61}
where the particle number densities are uncorrelated.

In fact, these distribution functions make the elastic collision terms
of the Boltzmann equation~(\ref{9.31}) equal to zero. In general,
however, the reactive collision terms~(\ref{9.35})~and (\ref{9.36}) do
not vanish, because the particle number densities in the distribution
functions~(\ref{9.61}) are not the equilibrium densities, which
characterize the chemical equilibrium and are related by the law of
mass action, Eq.~(\ref{a3}). In Eq.~(\ref{9.61}) we have introduced the
peculiar velocity $\xi_i^\a=c_i^\a-v_i$ where $v_i=\sum_{\a=1}^4
\varrho_\a v_i^\a\big/\sum_{\a=1}^4\varrho_\a$ is the bulk velocity of
the mixture.

From the Maxwellian distribution functions we can calculate the
forward reaction rate coefficient by substituting Eq.~(\ref{9.61}) in
the definition of the forward reaction rate coefficient,
Eq.~(\ref{9.44}). Substitution of the differential cross
section~(\ref{9.58a}) and integration of the
resulting equation leads to the result \be
\kk_f^{(0)}={(m_1m_2)^{3\over2}\over (2\pi kT)^3}\int
\exp\left[-{(m_{1}+m_{2})G_{12}^2\over 2kT} -{m_{12}g_{21}^2\over 2kT
  }\right] \sf\d_{12}^2\left[1-{2\epsilon_f\over
    {m_{12}\left(g_{21}\cos\theta\right)^2}} \right] g_{21}
\sin\theta\cos\theta \,\mathrm{d}\theta \,\mathrm{d}\varepsilon
\,\mathrm{d}{\bf g}_{21} \,\mathrm{d}{\bf G}_{12}, \ee{9.63}
since the solid-angle element is
$\mathrm{d}\Omega^\star=4\sin\theta\cos\theta \,\mathrm{d}\theta
\,\mathrm{d}\varepsilon$ where $\theta$ and $\varepsilon$ are the
angles characterizing the reactive collision process, while
$\bk_{21}\cdot\bg_{21}=g_{21}\cos\theta$. We have introduced the
relative velocity ${\bf g}_{21}$ and the center of mass velocity ${\bf
G}_{12}$ defined by the relation \be
g_i^{\a\b}=\xi_i^\b-\xi_i^\a,\qquad
G_i^{\a\b}=\frac{m_\a\xi_i^\a+m_\b\xi_i^\b}{m_\a+m_\b}.  \ee{9.63a}

The range of the integrals in the variables $G_{12}$ and $\varepsilon$
on the right-hand side of Eq.~\eqref{9.63} are $0\leq G_{12}<\infty$
and $0\leq\varepsilon\leq2\pi$, respectively. The range of the
integral over the angle $\theta$ follows from
Eq.~(\ref{9.58a}), according to which the normal
relative velocity has a lower bound so that the interval of
integration of $\theta$ is given by $0\leq\theta\leq \arccos
\sqrt{2\epsilon_f/m_{12}g_{21}^2}$. The range of the relative velocity
is $\sqrt{2\epsilon_f/m_{12}}\leq g_{12}<\infty$.

From the integration of (\ref{9.63}) we obtain the following
expression for the forward reaction rate coefficient
 \be \kk_f^{(0)}=\underline{\sqrt{\frac{8\pi
 kT}{m_{12}}}\,\sf^2\,\d_{12}^2\,e^{-\vf}}\,\left(1-\vf\,\Ef\,e^\vf\right),
\ee{9.64} a modified Arrhenius equation. Here,
$\vf$ is the forward activation energy in units of the thermal energy
$kT$, and $\Ef$ represents the exponential integral
$\Ef=\int_\vf^\infty\frac{e^{-y}dy}{y}$.

The underlined term in Eq.~(\ref{9.64}) is the usual Arrhenius
equation, which follows from substituting ${m_{12}g_{21}^2}$ for
${m_{12}\left(\bk_{21}\cdot\bg_{21}\right)^2}$ in the differential
cross section.  Inspection of Eq.~(\ref{9.64}) shows that the
reaction-rate coefficient is smaller for the latter case, in which a
reaction occurs only when the relative translational energy in the
direction of the line which joins the molecular centers exceeds
the activation energy. In the former case even a grazing collision
with relative translational energy larger than the activation energy
would lead to a chemical reaction.

\section{Slow Chemical Reactions}
\label{sec:5}
When dealing with chemical reactions within the framework of
the Boltzmann equation, we can analyze two regimes:
\begin{description}
\item[Fast reactions] the last stage of a chemical reaction. The
  system is nearby a chemical equilibrium state. The frequencies of
  the elastic and reacting collisions are of the same order of
  magnitude and the affinity is small in comparison with the thermal
  energy of the mixture, i.e., $\mathcal{A}\ll kT$;
\item[Slow reactions] the initial stage of a chemical reaction.  The
  system is far from chemical equilibrium. The elastic collisions are
  much more frequent than the reactive ones and the affinity is much
  larger than the thermal energy of the mixture, i.e., $\mathcal{A}\gg
  kT$.
\end{description}

In this work we shall analyze slow reactions. We write
the Boltzmann equations in the form
\begin{subequations}
  \begin{align}\label{6.1}
    \mathcal{D}f_\a +\lambda
    \xi_{i}^{\alpha}\frac{\partial{f}_\alpha}{\partial{x}_i}-
    \int\left[f_3 f_4\left( \frac{m_{12}}{m_{34}}\right)^3-f_1
      f_2\right]\sigma_{12}^\star g_{21}d\Omega^\star d{\bf
      c}_{\gamma} =\frac{1}{\lambda}
    \sum_{\beta=1}^4\int\left(f^\prime_\alpha f^\prime_\beta-f_\alpha
      f_\beta\right) g_{\beta\alpha}\sigma_{\beta\alpha}
    d\Omega\,d{\mathbf{c_\beta}},\\\label{6.2} \mathcal{D}f_\a
    +\lambda \xi_{i}^{\alpha}\frac{\partial{f}_\alpha}{\partial{x}_i}-
    \int\left[f_1 f_2\left(\frac{m_{34}}{m_{12}}\right)^3-f_3
      f_4\right]\sigma_{34}^\star g_{43} d\Omega^\star d{\bc}_{\gamma}
    = \frac{1}{\lambda}\sum_{\beta=1}^4\int\left(f^\prime_\alpha
      f^\prime_\beta-f_\alpha f_\beta\right)
    g_{\beta\alpha}\sigma_{\beta\alpha} d\Omega\,d{\mathbf{c_\beta}},
  \end{align}
\end{subequations}
where $\mathcal{D}=\partial/\partial t+v_i\partial/\partial x_i$ is
the material time derivative, and $\lambda$ a parameter of the order
of the Knudsen number. In Eq.~(\ref{6.1}) $\a=1,2$ corresponds to
$\gamma=2,1$, while in Eq.~(\ref{6.2}) $\a=3,4$, to
$\gamma=4,3$. Since we assumed the elastic collisions to be more
frequent than the reactive ones, Eqs.~(\ref{6.1})~and (\ref{6.2})
indicate that the reactive collision terms and the material time
derivatives are of the same order of magnitude, while the gradients of
the particle number densities, velocity and temperature of the mixture
are of the next order.

To obtain the distribution functions from Eqs.~(\ref{6.1})~and
(\ref{6.2}) we apply the Chapman-Enskog methodology. To that end,
we expand the material time derivative in a power series of a
parameter $\lambda$, i.~e., we write
 \be
  {\cal D}={\cal D}_0+\lambda{\cal D}_1+\lambda^2{\cal D}_2+\dots\,,
 \ee{9.125}along with the distribution functions \be
 f_\a=f_\a^{(0)}+\lambda f_\a^{(1)}+\lambda^2
 f_\a^{(2)}+\dots\,,\qquad \a=1,\dots,4.  \ee{9.126}
The parameter $\lambda$ in the distribution function will be set equal
to the unit later.

 We insert the expansions~(\ref{9.125})~and (\ref{9.126}) into the
 Boltzmann equations~(\ref{6.1})~and (\ref{6.2}). Comparison of equal
 powers of $\lambda$ leads to the following system of integral equations:
\begin{align}\label{9.127}
  0=&\sum_{\beta=1}^4\int\left(f^{(0)\prime}_\alpha
    f^{(0)\prime}_\beta-f^{(0)}_\alpha f^{(0)}_\beta\right)
  g_{\beta\alpha}\sigma_{\beta\alpha} d\Omega\,d{\mathbf{c_\beta}},
  \\\nonumber {\cal D}_0 f_\a^{(0)} -\int \left[f_3^{(0)}
    f_4^{(0)}\left( \frac{m_{12}}{m_{34}}\right)^3-f_1^{(0)}
    f_2^{(0)}\right]\sigma_{12}^\star g_{21}d\Omega^\star d{\bf
    c}_{\gamma}=&\sum_{\beta=1}^4\int\big[f^{(1)\prime}_\alpha
  f^{(0)\prime}_\beta+f^{(0)\prime}_\alpha f^{(1)\prime}_\beta
  \\\label{9.128} &-f^{(1)}_\alpha f^{(0)}_\beta -f^{(0)}_\alpha
  f^{(1)}_\beta\big] g_{\beta\alpha}\sigma_{\beta\alpha}
  d\Omega\,d{\mathbf{c_\beta}},\quad \a=1,2\quad \gamma=2,1,
  \\\nonumber {\cal D}_0 f_\a^{(0)} -\int \left[f_1^{(0)}
    f_2^{(0)}\left(\frac{m_{34}}{m_{12}}\right)^3-f_3^{(0)}
    f_4^{(0)}\right]\sigma_{34}^\star g_{43} d\Omega^\star
  d{\bc}_{\gamma}=&\sum_{\beta=1}^4\int\big[f^{(1)\prime}_\alpha
  f^{(0)\prime}_\beta+f^{(0)\prime}_\alpha f^{(1)\prime}_\beta
  \\\label{9.128a} &-f^{(1)}_\alpha f^{(0)}_\beta -f^{(0)}_\alpha
  f^{(1)}_\beta\big] g_{\beta\alpha}\sigma_{\beta\alpha}
  d\Omega\,d{\mathbf{c_\beta}},\quad \a=3,4\quad
  \gamma=4,3.  \end{align} Here we have only considered the integral
equations to first order in $\lambda$, i.~e., associated with the
zeroth order term $f^{(0)}_\a$ and the first-order correction $f^{(1)}_\a$.

The solution of the integral equation (\ref{9.127}) is the Maxwellian
distribution function (\ref{9.61}). To solve the integral
equations~(\ref{9.128})~and (\ref{9.128a}) we assume that the reaction
heat affects the Maxwellian distribution functions, so that
$f^{(1)}_\a$ can be expressed in terms of Sonine polynomials in the
peculiar velocity $\xi_i^\a$, i.~e.,  \be
f^{(1)}_\a=f_\a^{(0)}\sum_{n=1}^\infty \aa_{n\a}
\,S_{\frac{1}{2}}^{(n)}\left(\frac{m_\a\xi_\a^2}{2kT}\right),
\qquad\hbox{where}\qquad
S_{\frac{1}{2}}^{(n)}(x)=\sum_{k=0}^n\frac{\Gamma(n+3/2)}{k!(n-k)!\Gamma(k+3/2)}(-x)^k.
\ee{6.3}

The coefficients $\aa_{n\a}$ are constants to be determined
from the integral equations.  Up to second order terms in the
expansion, the distribution function can be written as
\be f_\a=f_\a^{(0)}\left[1+\aa_{1\a}
\left(\frac{3}{2}-\frac{m_\a\xi_\a^2}{2kT}\right)+
\aa_{2\a}\left(\frac{15}{8}-\frac{5m_\a\xi_\a^2}{4kT}
+\frac{m_\a^2\xi_\a^4}{8(kT)^2}\right)\right],
\qquad \a=1,\dots,4.  \ee{6.4}

Insertion of the distribution function~(\ref{6.4}) into the
definition~(\ref{temp}) of the temperature of constituent $\a$,
shows that all coefficients $\aa_{1\a}$ are zero. To determine
the remaining coefficients $\aa_{2\a}$ we multiply
Eqs.~(\ref{9.128})~and (\ref{9.128a}) by the Sonine polynomial
$[15/8-{5m_\a\xi_\a^2}/{4kT}+{m_\a^2\xi_\a^4}/{8(kT)^2}]$ and
integrate over all $d\bc_\a$, which shows that
\begin{align}\nonumber
&&-\frac{\left(1-e^{-\A}\right)\sf^2\d_{12}^2e^{-\vf}\left[1-4\,\vf+3\,\vf\,\Ef\,
e^{\vf}\right]{m_{12}^{\frac{3}{2}}}n_1 n_2}{8m_\a^2} \\\label{6.5}
&&=\sum_{\b=1}^4 \frac{n_\a n_\b
\sqrt{m_{\a\b}}}{(m_\a+m_\b)^2}\d_{\a\b}^2\left\{\frac{10m_\a^2+8m_\a
m_\b+13m_\b^2}{m_\a+ m_\b}\aa_{2\a}-15m_{\a\b}\aa_{2\b}\right\},\qquad
\a=1,2; \\\nonumber
&&\frac{\left(1-e^{-\A}\right)\sf^2\d_{12}^2e^{-\vf}\left\{1-4\,\vf+3\,\vf\,\Ef\,
e^{\vf}-4\q\left[1+\q-\vf\,\Ef\,e^{\vf}\left(3+\q\right)\right]\right\}m_{34}^5n_1
n_2}{8m_\a^2m_{12}^{\frac{7}{2}}} \\\label{6.6} &&=\sum_{\b=1}^4
\frac{n_\a n_\b
\sqrt{m_{\a\b}}}{(m_\a+m_\b)^2}\d_{\a\b}^2\left\{\frac{10m_\a^2+8m_\a
m_\b+13m_\b^2}{m_\a+ m_\b}\aa_{2\a}-15m_{\a\b}\aa_{2\b}\right\},\qquad
\a=3,4.\end{align}

From the algebraic system of Eqs.~(\ref{6.5})~and
(\ref{6.6}) we can determine the four coefficients $\aa_{21},\dots,
\aa_{24}$ of the distribution functions (\ref{6.4}). The expressions
for these coefficients are too long to be presented here.

To determine the forward  and reverse reaction rate coefficients, we
insert each distribution function~(\ref{6.3})
in the corresponding definition, in Eq.~(\ref{9.44}). Integration of
the resulting equalities then yields the expressions
 \begin{align}\label{6.7a} \kk_f=\kk_f^{(0)}
\left\{1-\underline{\frac{1-4\vf+3\vf\,\Ef\,e^\vf}{1-\vf\,\Ef\,e^\vf}\,\frac{\aa_{21}m_2^2+\aa_{22}m_1^2}{8(m_1+m_2)^2}}\right\},\end{align}
\begin{align}\label{6.7b}
\kk_r=\left(\frac{m_1m_2}{m_3m_4}\right)^{\frac{3}{2}}e^{\q}\kk_f^{(0)}\left\{1-
\underline{\frac{1-4\left(\vf+\q+{\q}^2\right)+\vf\,\Ef\,e^\vf\left(3+12\q+4{\q}^2\right)}{1-\vf\,\Ef\,e^\vf}\frac{\aa_{23}m_4^2+\aa_{24}m_3^2}{8(m_3+m_4)^2}}\,\right\}.
\end{align}

The underlined terms are corrections to the forward  and
reverse reaction rate coefficients, for non-Maxwellian distribution
functions. The corrections depend on the coefficients $\aa_{21},\dots,
\aa_{24}$ of the distribution functions (\ref{6.4}), on the reaction
heat, and on the activation energy of the forward reaction.

\section{Reaction rates for $H_2 +Cl\rightleftharpoons HCl+H$}
\label{sec:6}
We now apply the results of Secs.~\ref{sec:4}~and \ref{sec:5} to evaluate
the rate coefficients for the reaction $H_2
+Cl\rightleftharpoons HCl+H$. To that end we need to know
certain characteristic parameters for the constituents of the mixture,
such as masses, diameters, forward and reverse activation energies,
reaction heat and steric factors. Table 1 lists the
molecular weights $M_\alpha$ and the coefficients of shear
viscosity $\mu_\alpha$ at temperature $T=293$~K
for the single constituents $H$, $H_2$, $Cl$ and $HCl$ \cite{LB1}.
From the expression of the coefficient of shear viscosity for
hard-sphere molecules (see e.g. \cite{GK})
 \be \mu_\alpha={5\over16 \d_\alpha^2}\sqrt{m_\a kT\over\pi},
 \ee{9.103}
we can calculate the diameters of the single constituents $H_2$ and
 $HCl$. For the constituent $Cl$ we take the diameter to be twice
 the atomic radius, whereas for the constituent $H$ we take the diameter
 to be of order of twice the Bohr radius $a_0=0.529\times 10^{-10}$~m.

\begin{center}
\begin {table}[h] \centering\footnotesize{
\begin{tabular}{|c|c|c|c|c|} \hline Gas & H& H$_2$& Cl & HCl \\ \hline
$M_\alpha$&1.008&2.016&35.453&36.461\\ $\mu_\alpha$ $(\times10^{-5}$
Pa s)& -- &0.841& --&1.332\\ $d_\alpha$ $(\times10^{-10}$
m)&1.06&2.78&1.99&4.55\\ \hline
\end{tabular}}
\caption{Molecular weights, viscosities and molecular diameters.}
\end{table}
\end{center}

Table 2 lists the coefficient $A$ of the Arrhenius equation
$\kk^{(0)}_f=Ae^{-\epsilon_f^\star}$, the forward, and the reverse
activation energies \cite{LB2}.  The reference temperature for this
reaction is 300~K and the reaction heat was obtained from the
relation $Q=\epsilon_f-\epsilon_r$. The steric factors in this
table were calculated from the identification $A=\sqrt{8\pi
    kT/m_{12}}\,\sf^2\,\d_{12}^2$, from Eq.~(\ref{9.64}).

\begin{center}
\begin {table}[h] \centering
\begin{tabular}{|c|c|c|c|c|c|} \hline Reaction & $A$(m$^3$/mol
s)&$\epsilon_f$(kJ/mol)&$\epsilon_r$(kJ/mol)&$Q$(kJ/mol)&$\s_f$\\
\hline
 $H_2+Cl \rightharpoonup HCl+H$&7.94$\times 10^7$
&23.03&18.84&4.19&0.648\\ $HCl+H \rightharpoonup H_2+Cl$&4.68$\times
10^7$ &18.84&23.03&-4.19&0.497\\
   \hline
\end{tabular}
\caption{Arrhenius coefficients, forward and reverse activation
energies, reaction heat and steric factors.}
\end{table}
\end{center}

To analyze a slow reaction, in which the concentrations of the
reagents is larger than those of the products, we start out with a
preliminary evaluation of the factor
$\left(\frac{m_1m_2}{m_3m_4}\right)^{\frac{3}{2}}e^{\q}$ appearing in
the reverse reaction rate coefficient~(\ref{6.7b}). If we adopt the
tabulated values and consider the reaction in the direction
$H_2+Cl\rightharpoonup HCl+H$, a very large ratio results between the reverse  and the
forward reaction coefficients. We therefore discard this
alternative and consider only the reaction in the direction
$HCl+H\rightharpoonup H_2+Cl$, for which the ratio is small.

To determine the coefficients $\aa_{2\a}$ from the
algebraic system of Eqs.~(\ref{6.5})~and (\ref{6.6}) we need
to know the temperature and the concentrations of the
constituents $x_\a=n_\a/n$ where $n=\sum_{\b=1}^4 n_\b$ and
$\sum_{\a=1}^4 x_\a=1$. Here we analyze the case $x_1=x_2$
and $x_3=x_4$. Figure~1 shows the coefficients $\aa_{2\a}$, plotted as functions
of the temperature in the range 400~K $\leq T\leq$ 1200~K
for two concentrations: $x_1=0.30$ and $x_1=0.35$. The figure shows
that the deviations from the Maxwellian distribution
functions\textemdash which are given by the coefficients $\aa_{2\a}$\textemdash
increase with the temperature. In comparison
with the departure for the constituent $H_2$, the departures for the
constituents $HCl$, $H$ and $Cl$ are not very large.

Figure 2 shows the dependence of the reaction rate coefficients on the
temperature in the range 400~K $\leq T\leq$ 1200~K, for two
concentrations: $x_1=0.30$ and $x_1=0.35$. The left (right) panel represents
the forward reaction (reverse reaction) rate coefficient. Consider
first the influence of the collision geometry on the reaction rate
coefficients. The dashed-dotted lines in the inset left panel and in
the right panel represent the Arrhenius law, given by the
underlined term in Eq.~(\ref{9.64}), while the solid lines correspond to
the modified Arrhenius equation~(\ref{9.64}), when the geometry of the
collision is taken into account, i.e., when a reaction occurs only if
the relative translational energy in the direction of the line
joining the molecular centers exceeds the activation
energy. The reaction rates in the former case are larger
than those in the latter, an indication that more reactions occur in the
former case,  due to the fact than even grazing collisions
with translational energy larger than the activation energy lead to
chemical reactions.  In the same figures we also plot the influence
of the departures from the Maxwellian distribution functions on the
reaction rate coefficients. The left and the right
panels show that the non-Maxwellian distribution functions reduce the
reaction rate coefficients and that the forward reaction rate
coefficient is less reduced than the reverse reaction one. The
reduction becomes more pronounced as the reagent concentration grows.

\begin{figure}
\begin{center}
\includegraphics[width=7cm]{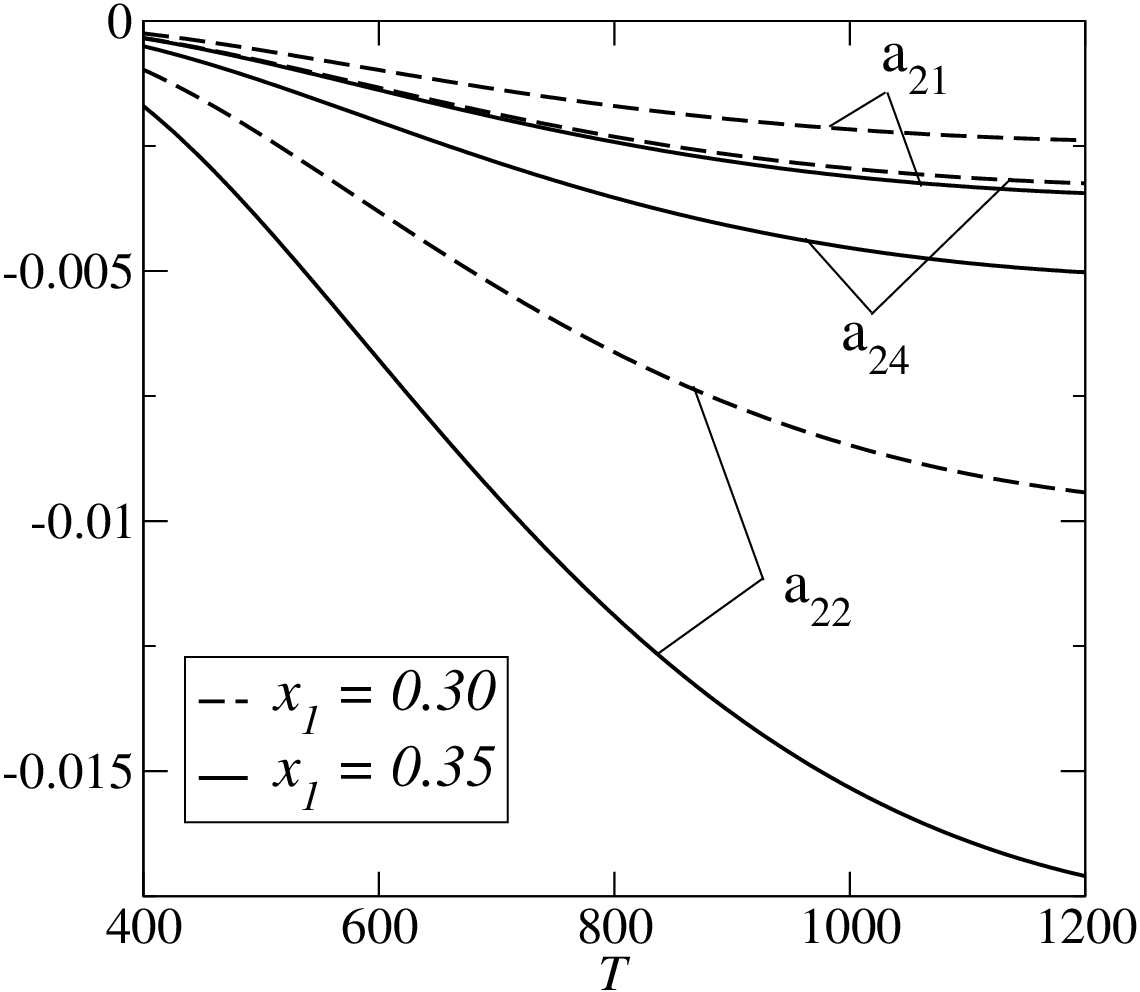}\hskip0.1cm
\includegraphics[width=7cm]{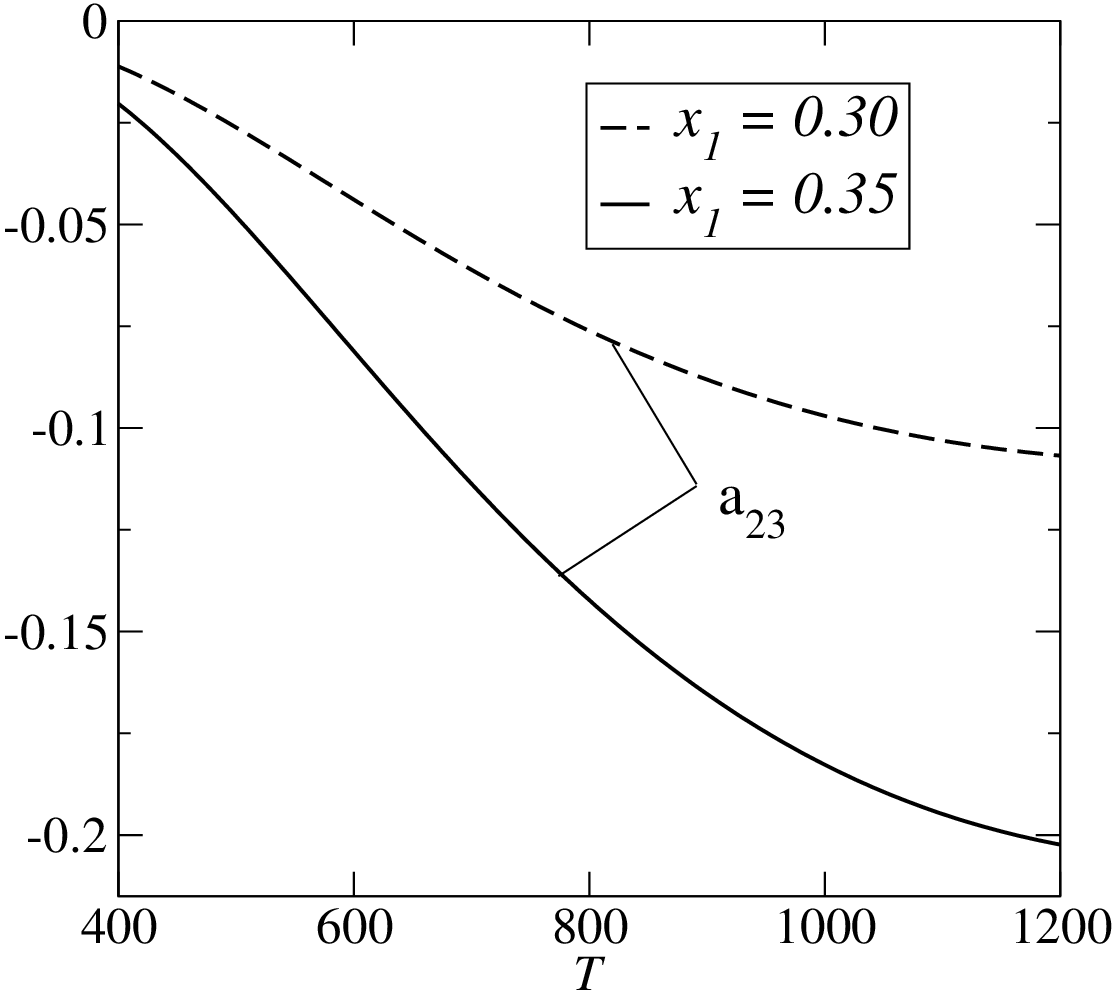}
\caption{Coefficients $\aa_{2\a}$ as functions of the temperature for
two concentrations: 0.30 and 0.35.}
\end{center}
\end{figure}

\begin{figure}
\begin{center}\vskip0.5cm
\includegraphics[width=7cm]{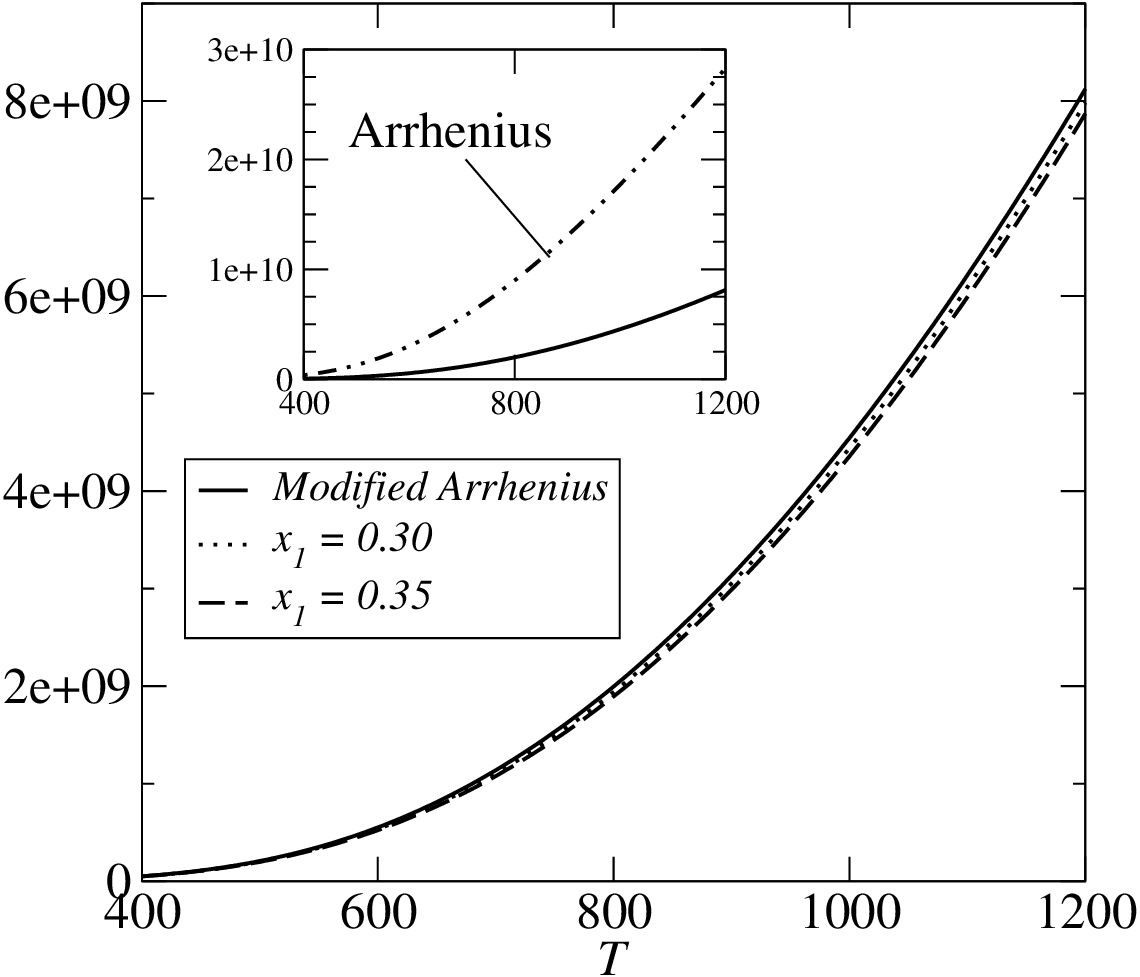}\hskip0.1cm
\includegraphics[width=7cm]{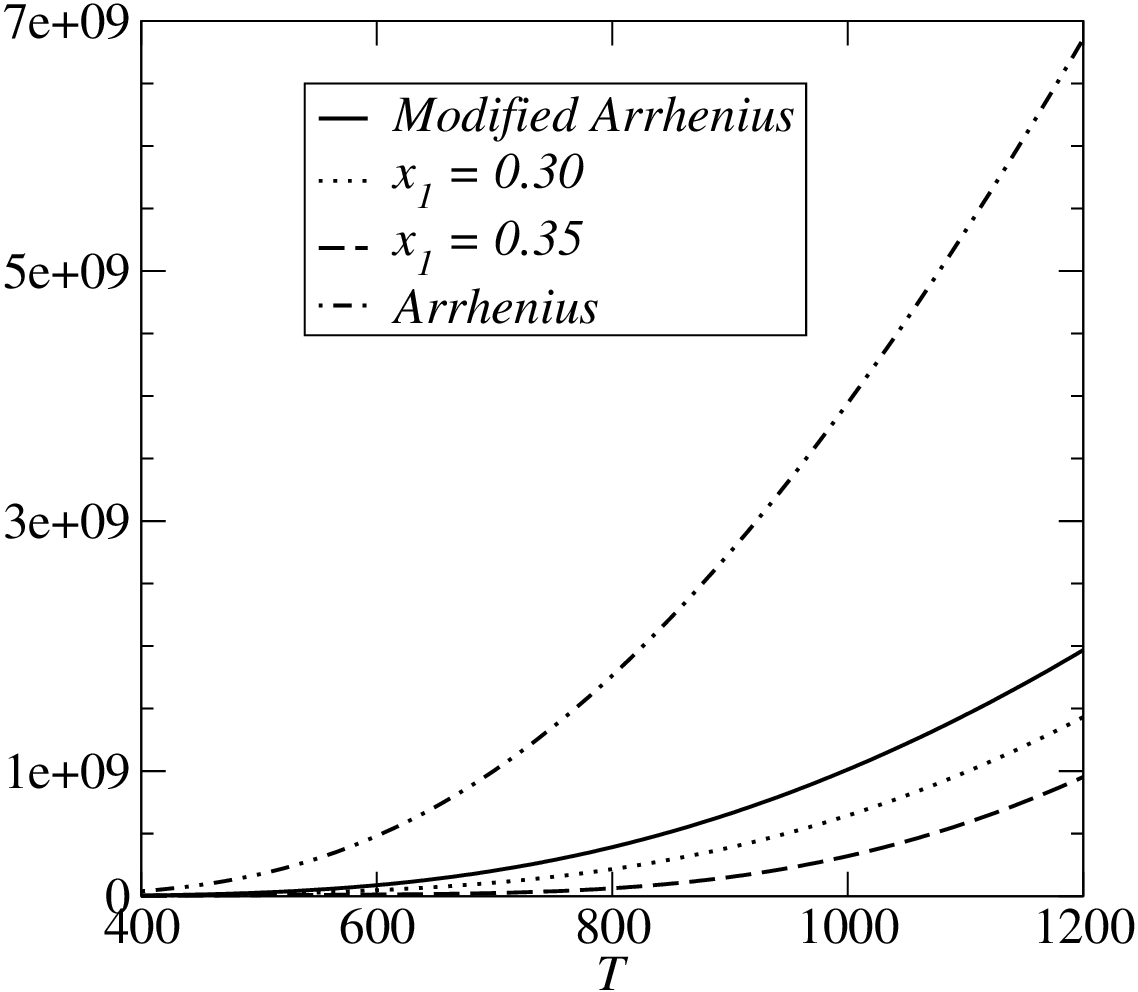}
\caption{Reaction rate coefficients in dm$^3$/(mol s) as functions of
temperature in K. Left panel: forward reaction rate; right panel:
reverse reaction rate.}
\end{center}
\end{figure}

Once the forward and reverse reaction rates are known, we can compute
the entropy production rate. According to a phenomenological
theory (see e.g. \cite{Pri,dGM}) the entropy-density rate $\varsigma$
is given by \begin{align}\label{za}
\varsigma=\frac{\mathcal{A}}{T}\frac{d\xi}{dt},\qquad
\hbox{where}\qquad \frac{d\xi}{dt}=\kk_f n_1n_2-\kk_r n_3n_4, \end{align}
the so-called reaction velocity.

Equation~\eqref{za} can also be deduced from the Boltzmann equation;
for more details see Ref.~\onlinecite{3,GK}.  In view of the law of
mass action $\kk_f n_1^{\rm eq}n_2^{\rm eq}=\kk_r n_3^{\rm eq}n_4^{\rm
  eq}$, which shows that, in equilibrium, the forward and the reverse
reactions occur with the same probability, the affinity~(\ref{a4})
becomes \begin{align} \mathcal{A}=kT\ln\left(\frac{\kk_f n_1n_2}{\kk_r
      n_3n_4}\right), \end{align} so that the entropy production
rate~(\ref{za}) can be written as
\begin{align}\label{za1} \frac{\varsigma}{n^2}=k\ln\left(\frac{\kk_f
x_1x_2}{\kk_r x_3x_4}\right)\left(\kk_f x_1x_2-\kk_r x_3x_4\right).
\end{align}

As one would expect, the entropy production rate is positive
semi-definite\textemdash thanks to the inequality $\left(x-1\right)\ln x\geq0$
which is valid for all $x>0$, the equality being valid for $x=1$.

Figure 3 shows the entropy production rate~(\ref{za1}) as a function
of the temperature for the Arrhenius and modified Arrhenius cases, to
show that in the former case the rate is larger than in the
latter. This is expected, given the entropy production
rate dependence on the reaction rates. For larger concentrations
$x_1$ the entropy production rate increases in both cases.

\begin{figure}
\begin{center}\vskip0.5cm
\includegraphics[width=7cm]{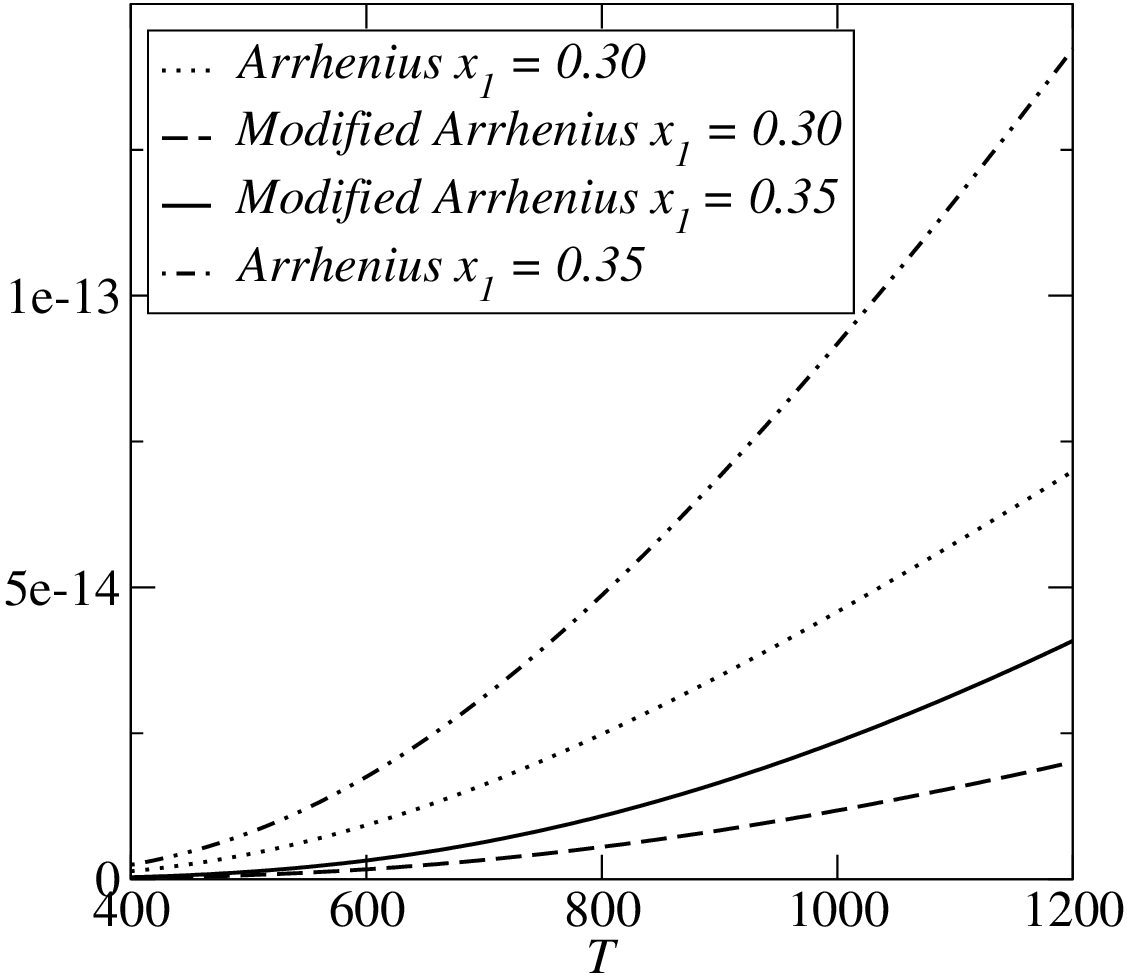}
\caption{Entropy production rate in J~dm$^3$/(K\;mol\;s) as function
of the temperature in K.}
\end{center}
\end{figure}

\section{Conclusions}

We have used the Boltzmann equation to analyze a bimolecular chemical
reaction in its initial stage. As an illustration, we computed the
reaction rates and the entropy production rate for the typical
bimolecular reaction $H_2 +Cl\rightleftharpoons HCl+H$.  To solve the
Boltzmann equation by the Chapman-Enskog method we have expanded the
distribution function in Sonine polynomials up to the second
order. Our reactive differential cross section allows a reaction to
occur only when the relative translational energy in the direction of
the line of centers which joins the centers of the molecules is
greater than the activation energy. With this cross section the
calculated reaction rates are substantially smaller than those
obtained from the line-of-centers model, which requires the relative
translational energy to be larger than the activation energy. We
showed that the restriction to large relative translational energies
along the line of centers affects the calculated rates more than the
departure from the Maxwellian distribution function.

\section*{Acknowledgements} GMK acknowledges the support by CNPq and
TGS acknowledges the support by CAPES.

\end{document}